\documentclass[pra,showpacs,showkeys,twocolumn,amsmath,amssymb,superscriptaddress,preprintnumbers,floatfix]{revtex4-1}
\usepackage{bm,amsfonts}
\usepackage{graphicx,color, wasysym}
\usepackage{textcomp}

\begin{document}
\renewcommand{\arraystretch}{2.3}

\title{L\'{e}vy stable two-sided distributions: exact and explicit densities for asymmetric case}
\author{K. G\'{o}rska}
\email{kasia\_gorska@o2.pl}

\author{K. A. Penson}
\email{penson@lptl.jussieu.fr}

\affiliation{Laboratoire de Physique Th\'{e}orique de la Mati\`{e}re Condens\'{e}e (LPTMC),
Universit\'{e} Pierre et Marie Curie, CNRS UMR 7600,
Tour 13 - 5i\`{e}me \'et., Bo\^{i}te Courrier 121, 4 place Jussieu, F 75252 Paris Cedex 05, France}

\pacs{05.40.Fb, 05.40.-a, 02.50.Ng}

\begin{abstract}
We study the one-dimensional L\'{e}vy stable density distributions $g(\alpha, \beta; x)$ for $-\infty < x < \infty$, for rational values of index $\alpha$ and the asymmetry parameter $\beta$: $\alpha = l/k$ and $\beta = (l - 2r)/k$, where $l, k$ and $r$ are positive integers such that $0 < l/k < 1$ for $0 \leq r \leq l$ and $1 < l/k \leq 2$ for $l-k \leq r \leq k$. We treat both symmetric ($\beta = 0$) and asymmetric ($\beta \neq 0$) cases. We furnish exact and explicit forms of $g(\alpha, \beta; x)$ in terms of known functions for any admissible values of $\alpha$ and $\beta$ specified by a triple of integers $k$, $l$ and $r$. We reproduce all the previously known exact results and we study analytically and graphically many new examples. We point out instances of experimental and statistical data that could be described by our solutions.
\end{abstract}

\maketitle


The probability distributions characterizing anomalous diffusive behaviour have been a subject of intense activity on experimental and theoretical side. Among various forms proposed, that of the heavy-tailed L\'{e}vy stable laws is of widespread use, mostly due to their universal presence in such diversified fields as econophysics \cite{EScalas07, SBorak05}, physics of amorphous materials \cite{RKutner99}, geology \cite{Che-YiYang09}, biophysics \cite{IMSololov10}, statistics of phone networks \cite{CAHidalgo08}, internet traffic \cite{GyTerdik09} and dynamics of human contacts \cite{MPFreeman10}. Entire monographs are devoted to the thorough study of this huge field \cite{VVUchaikin99}. Comprehensive reviews are available reporting on the state of the art \cite{APiryatnska05, WAWoyczynski01, RMetzler09, BDybiec10}. Further references may be traced back from \cite{NKorabel10, KAPenson10}.

The goal of the paper is to give exact and explicit expression for the two-sided  L\'{e}vy stable distributions $g(\alpha, \beta; x)$, $0 < \alpha \leq 2$, for symmetric ($\beta = 0$) and asymmetric ($\beta \neq 0$) cases. Since $g(1, \beta; x)$ requires special treatment \cite{VVUchaikin99} we omit the case $\alpha = 1$.

The probability density function (PDF), $g(\alpha, \beta; x)$, is called stable if the product of characteristic functions (CF) of two such laws is a CF of another law of the same type. The general PDF of this type $g(\alpha, \beta; x)$, where  either $-\infty < x < \infty$, or $x$ is confined to one of semiaxes, see below, has the CF defined as the Fourier transform in the form \cite{VVUchaikin99, HBergstrom52, WFeller70, IMSokolov00}
\begin{equation}\label{eq1}
\hat{g}(\alpha, \beta; p) = \exp\left[- |p|^{\alpha} \exp\left(i \beta \pi\, sgn(p)/ 2\right)\right],
\end{equation} 
where $|p|$ and $sgn(p)$ are the absolute value and the sign of $p$, respectively, and  $\hat{g}(\alpha, \beta; p)$ satisfy the relation $\hat{g}(\alpha, \beta; -p) \,=\, \hat{g}(\alpha, -\beta; p)$. According to the values of parameters $\alpha$ and $\beta$ we can distinguish the following variants: (i) for $0 < \alpha < 1$ and $|\beta| \leq \alpha$: we have for $\beta = -\alpha$  one-sided PDFs defined for $0 \leq x < \infty$, whereas for $\beta = \alpha$ they are defined only for $-\infty < x \leq 0$, otherwise they are two-sided, i. e. $-\infty < x < \infty$; (ii) for $1 < \alpha \leq 2$ and $|\beta| \leq 2 - \alpha$ the PDFs are always two-sided. Only under these restrictions on $\alpha$ and $\beta$ the positivity of $g(\alpha, \beta; x)$ for all allowed $x$ is guaranteed \cite{VVUchaikin99, WFeller70, IMSokolov00}.  

The finding of exact and explicit form of two-sided $g(\alpha, \beta; x)$ turned out to be a true challenge. In the literature we can find only a limited number of exact formulae for $g(\alpha, \beta; x)$. The two-sided asymmetric cases include $\alpha = 1/3$ and $\beta = \pm 2/3$ \cite{VVUchaikin99}, $\alpha = 3/2$ and $\beta = \pm 1/2$ \cite{VVUchaikin99, WRSchneider86}. The symmetric cases ($\beta = 0$) concern  $\alpha = 4/3$ \cite{TMGaroni02, AHatzinikitaz08},  $\alpha = 5/4$ and $6/5$ \cite{AHatzinikitaz08}, $\alpha = 1/3$ and $1/2$ \cite{TMGaroni02, AHatzinikitaz08}, $\alpha = 1/5$, $1/4$, $2/5$, $3/5$, $3/4$, and $4/5$ \cite{AHatzinikitaz08}. The exact solutions for one-sided case $\beta = -\alpha$ for rational $\alpha$ have been recently obtained in \cite{KAPenson10}. For any other values of $\alpha$ and $\beta$ for two-sided situation the only source of information are numerical calculations, often problematic if not impossible for small values of $\alpha$ \cite{JPNolan99, RHRimmer05}.

In what follows we shall indicate how the approach of \cite{KAPenson10} can be extended to obtain new exact representations for two-sided case with rational values of $\alpha$ and $\beta$. One should be reminded at this point that the functional structure of $g(\alpha, \beta; x)$ depends on an essential way on the value of $\alpha$ \cite{WFeller70, VVUchaikin99}: for $1 < \alpha \leq 2\,$ $g(\alpha, \beta; x)$ is a unique function for both signs of $x$. On the contrary, for $0 < \alpha < 1$, $\beta \neq 0$, $-\alpha$, the function $g(\alpha, \beta; x)$ is obtained by matching of two different functions $g(\alpha, -\beta; x)$ for $x < 0$ with $g(\alpha, \beta; x)$ for $x > 0$, at the point $x = 0$. In any case it is sufficient to use $g(\alpha, \beta; x)$ defined as \cite{WFeller70, IMSokolov00, BDHughes95}
\begin{equation}\label{eq2}
g(\alpha, \beta; x) \,=\, \frac{1}{\pi}\, \Re \int_{0}^{\infty}dp\, \, e^{-i p x} e^{- p^{\alpha} \exp(i \beta\, \pi/2)}.
\end{equation}  
Before going to the most general case let us embark upon an intermediate situation where for a given $\alpha$ only certain values of $\beta$ intervene.

\begin{figure}[h]
\begin{center}
\includegraphics[scale=0.4]{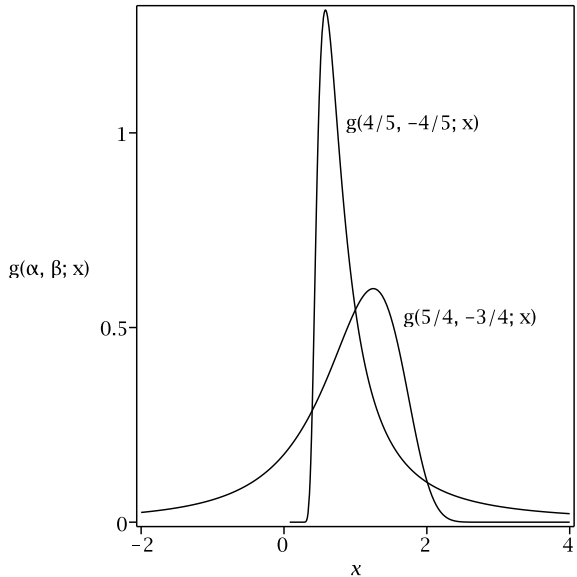}
\caption{\label{fig1} Comparison of the one-sided distribution $g(4/5, -4/5; x)$ and its dual two-sided counterpart $g(5/4, -3/4; x)$, see Eq.~(\ref{eq3}).}
\end{center}
\end{figure}

The general form of one-sided $g(\alpha, -\alpha; x)$ for rational $\alpha = l/k$, $l < k$, where $l, k$ are integers, was recently presented in \cite{KAPenson10} where it is denoted by $g_{\alpha}(x)$. As an initial approach to the two-sided case we shall generate \textit{certain} two-sided solutions $g(\alpha^{\star}, \beta^{\star}; x)$, $\alpha^{\star} = \alpha^{-1}$, $1 < \alpha^{\star} \leq 2$ and $\beta^{\star} = \alpha^{\star} - 2$, via the duality law (DL) \cite{VVUchaikin99, WFeller70} applied to one-sided $g(\alpha, -\alpha; x)$. The DL implies, for $\alpha < 1$, $g(\alpha^{\star}, \alpha^{\star}-2; x) = x^{-(1 + 1/\alpha)} g(\alpha, -\alpha; x^{-1/\alpha})$, from where for rational $\alpha = (\alpha^{\star})^{-1} = l/k$ we have, according to \cite{KAPenson10}
\begin{eqnarray}\nonumber
\lefteqn{g(\alpha^{\star}, \alpha^{\star} - 2; x) = } \\ \label{eq3}
& & \sum_{j = 1}^{k-1} \frac{b_{j}(k, l)}{x^{1-j}} \,_{l+1}F_{k} \left(^{1, \Delta(l, 1 + j l/k)} _{\Delta(k, j+1)} \Big\vert (-1)^{k-l} \frac{l^l\, x^k}{k^k} \right).
\end{eqnarray}
In Eq.~(\ref{eq3}) $_{p}F_{q}\left(^{(a_p)}_{(b_q)} \vert z\right)$ is the generalized hypergeometric function with the upper and lower parameter lists equal to $(a_{p})$ and $(b_{q})$ respectively \cite{APPrudnikov3}, and $\Delta(n, a) = \frac{a}{n}, \frac{a+1}{n}, \ldots, \frac{a + n-1}{n}$ is a special list of $n$ parameters. The numerical coefficients $b_{j}(l, k)$ in Eq.~(\ref{eq3}) have the form of Eq.~(4) in \cite{KAPenson10}. We emphasize that while the distributions $g(\alpha, -\alpha; x)$ are defined for $x \geq 0$, $g(\alpha^{\star}, \alpha^{\star} - 2; x)$ are automatically valid for $-\infty < x < \infty$. To carry out this procedure we consider Eq.~(6) of \cite{KAPenson10} for $p = 4$ which corresponds in our notation to $g(4/5, -4/5; x)$, $x \geq 0$. The DL yields $g(5/4, -3/4; x)$ for $-\infty < x < \infty$ whose exact form is given by Eq.~(\ref{eq3}) for $l=4$ and $k=5$. In view of previous remarks and the values of $\alpha^{\star}$ and $\beta^{\star}$ involved, $g(5/4, -3/4; x)$ is the unique function for both semiaxes. In Fig.~\ref{fig1} we compare both of these distributions. As far as we know the function $g(5/4, -3/4; x)$ is a new exact solution for two-sided case. According to Eq.~(\ref{eq3}) it can be represented, after appropriate simplifications of their parameter lists $(a_{p})$ and $(b_{q})$, as a sum of four hypergeometric functions of type $_{3}F_{3}$ of argument $\left(-4^4\, x^5/5^5\right)$. For a given $0 < \alpha < 1$ this DL route based on one-sided solutions in \cite{KAPenson10} will always yield special two-sided densities in the form $g\left(\frac{1}{\alpha}, \frac{1 - 2\alpha}{\alpha}; x\right)$, $-\infty < x < \infty$. They should be hitherto considered as known \cite{ref0}.

In fact, we have achieved a more ambitious goal by finding exact $g(\alpha, \beta; x)$ without restrictions on $\beta$ imposed by the DL. By extending the method of \cite{KAPenson10} we have established a general and universal formula for $g(\alpha, \beta; x)$ which encompasses both one- and two-sided cases. The general form of Eq.~(\ref{eq2}) for appropriate rational $\alpha = l/k$ and $\beta = \alpha - 2r/k$, where $l, k$ and $r$ are positive integers (see \cite{ref1}), is given by the exact expression:
\begin{eqnarray}\nonumber
\lefteqn{g(\alpha, \beta; x) = \sum_{j = 1}^{M-1}\, \frac{c_{j}(l, k, r)}{x^{1 \mp j\,l/M}} } \\ \label{eq4}
& & \times \,_{m+1}F_{M}\left(^{1, \Delta(m, 1 + j\, m/M)}_{\Delta(M, j+1)} \Big\vert (-1)^{r - M} \frac{m^m\, x^{\pm l}}{M^M} \right),
\end{eqnarray}
where $m = \min(l, k)$ , $M = \max(l, k)$, the lower sign is for $l < k$ and the upper sign for $l > k$, with the coefficients
\begin{eqnarray}\nonumber
c_j(l, k, r) &=& \frac{M^{1/2 - j} \, m^{1/2 + j m/M}}{2^{-r} \, (2\pi)^{(l+k)/2}} \,  \left[\prod_{i=0}^{m-1} \Gamma\left(\frac{j}{M} + \frac{i+1}{m}\right)\right] \\ \label{eq5}
& \times & \frac{\left[ \prod_{i = 1}^{j} \Gamma\left(\frac{i-j-1}{M}\right) \right]\, \left[ \prod_{i = j+2}^{M} \Gamma\left(\frac{i-j-1}{M}\right) \right]}{\prod_{i=0}^{r-1} \left[\sin\left(\pi\frac{i}{r} - \pi\frac{j}{M}\right)\right]^{-1}}.
\end{eqnarray}

Here is the sketch of derivation of Eqs.~(\ref{eq4}) and (\ref{eq5}).~They result from the application of the Mellin transform to $g(\alpha, \beta; x)$ of Eq.~(\ref{eq2}), $\mathcal{M}\left[g(\alpha, \beta; x); s\right] = \int^{\infty}_{0}dx\, x^{s-1} g(\alpha, \beta; x)$ for complex $s$, which is equal to $\frac{1}{\alpha\pi} \cos\left(\pi \frac{\beta}{2\alpha} + \pi\,s\frac{\alpha - \beta}{2\alpha}\right) \Gamma(s)\, \Gamma\left(\frac{1-s}{\alpha}\right)$. Then, $g(\alpha, \beta; x)$ will be perceived as the inverse Mellin transform,~i.~e.~it is formally equal to $\mathcal{M}^{-1}\left[\frac{1}{\alpha\pi} \cos\left(\pi\frac{\beta}{2\alpha} + \pi\,s\frac{\alpha - \beta}{2\alpha} \right) \Gamma(s) \Gamma\left(\frac{1-s}{\alpha}\right); x \right]$.~The~next steps involve the use of Euler's reflection formula for cosinus, the passage to rationals $\alpha = l/k$ and $\beta = (l - 2r)/k$ and the use of Gauss-Legendre multiplication formula for all gamma functions. Putting all the terms together, we employ, as an intermediate tool, the storing of the inverse Mellin transform as the Meijer~G function $G^{m, n}_{p, q}\left( z \vert^{\ldots}_{\ldots}\right)$ \cite{APPrudnikov3}. The final use of conversion formula 8.2.2.3, p.~618 of Ref.~\cite{APPrudnikov3} yields Eqs.~(\ref{eq4}) and (\ref{eq5}).

\begin{figure}[h]
\begin{center}
\includegraphics[scale=0.4]{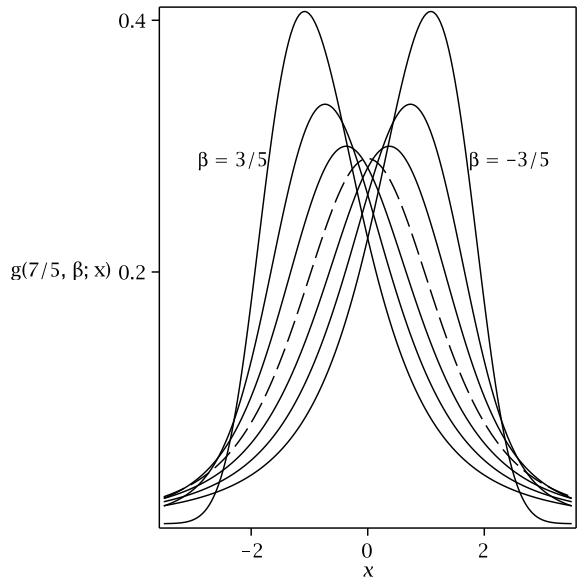}
\caption{\label{fig2} Comparison of asymmetric distributions $g(\alpha, \beta; x)$ for $\alpha \,=\, 7/5$ and $\beta = -3/5$, $-2/5$, $-1/5$, $1/5$, $2/5$, respectively, starting from the right. The dashed curve is the symmetric case $\alpha = 7/5$, $\beta = 0$.}
\end{center}
\end{figure}

\begin{figure}[h]
\begin{center}
\includegraphics[scale=0.4]{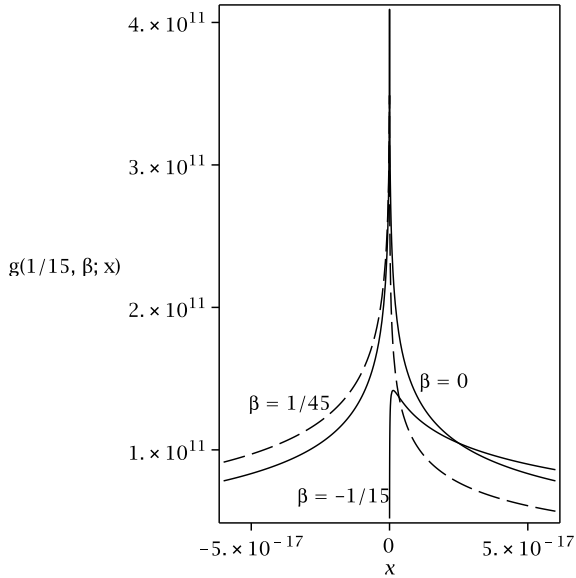}
\caption{\label{fig3} Comparison of $g(\alpha, \beta; x)$ for $\alpha \,=\, 1/15$ and $\beta = -1/15$, $0$, $1/45$. For $\beta = -1/15$ the distribution is one-sided. For $\beta = 1/45$ the exceedingly large peak is at very small $x < 0$.}
\end{center}
\end{figure}

\begin{figure}[h]
\begin{center}
\includegraphics[scale=0.4]{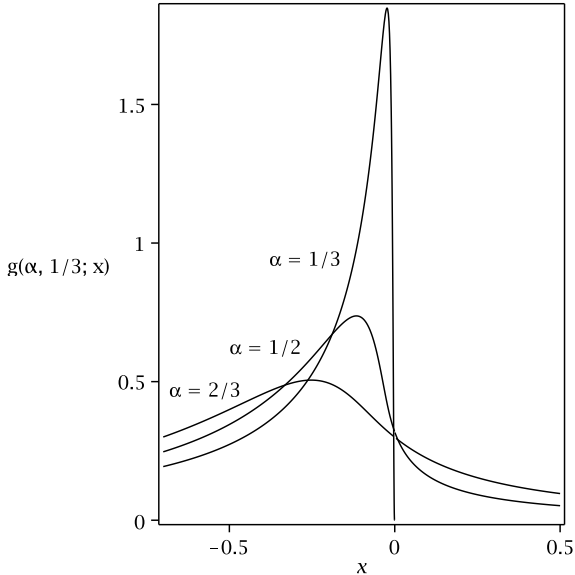}
\caption{\label{fig4} Comparison of asymmetric distributions $g(\alpha, \beta; x)$ for $\beta \,=\, 1/3$ and $\alpha = 1/3$, $1/2$ and $2/3$. The distribution $g(2/3, 1/3; x)$ was calculated with Eqs.~(\ref{eq6}) and (\ref{eq7}).}
\end{center}
\end{figure}

The actual construction of $g(\alpha, \beta; x)$ from Eqs.~(\ref{eq4}) and (\ref{eq5}) for a given triple $(l, k, r)$ boils down into three distinct alternatives: (a)~$\alpha = l/k < 1$, $\beta = \alpha - 2 r/k = -\alpha$ which gives $r = l$; it yields a one-sided $g\left(\frac{l}{k}, -\frac{l}{k}; x\right)$ for $x \geq 0$, elaborated in \cite{KAPenson10}; furthermore, for $\alpha = l/k < 1$, but for $\beta = \alpha - 2 r/k = \alpha$, which gives $r = 0$, it yields a one-sided $g\left(\frac{l}{k}, \frac{l}{k}; x\right)$ defined only for $x \leq 0$; (b)~for $\alpha = l/k$ such that $1 < l/k \leq 2$ and $|\beta| = |(l - 2r)/k| \leq 2 - l/k$, which implies $l-k \leq r \leq k$, both two-sided density functions $g\left(\frac{l}{k}, \frac{l-2r}{k}; x\right)$ and $g\left(\frac{l}{k}, \frac{2r-l}{k}; x\right)$ are defined on  $-\infty < x < \infty$ (and are mutually symmetric with respect to $(0, y)$, see Fig.~\ref{fig2}); (c) $\alpha = l/k < 1$, but $|\beta| = |(l - 2r)/k| < \alpha$, which implies $0 < r < l$; here the density $g(\alpha, \beta; x)$ decomposes into two \textit{different} functions according to the sign of $x$, and is given by $g\left(\frac{l}{k}, \frac{l-2r}{k}; x\right) \theta(-x) + g\left(\frac{l}{k}, \frac{2r-l}{k}; x\right) \theta(x)$, with $\theta(x)$ the Heaviside function. The matching at $x = 0$ of these two components assures the continuity of $g(\alpha, \beta; x)$ at $x = 0$, along with all its higher derivatives, see Figs.~\ref{fig2}, \ref{fig3} and \ref{fig4}.

Eqs.~(\ref{eq4}) and (\ref{eq5}) can be equivalently represented as a single infinite series derived by Bergstr\"{o}m and Feller \cite{HBergstrom52, WFeller70}, which is a two-sided variant of the Humbert expansion \cite{PHumbert45}. This formula (\textit{vide} Eqs.~(4) and (6) in \cite{HBergstrom52}) is however very slowly convergent in both regions of $\alpha \to 0$ and $\alpha \apprle 1$. On the contrary, the Eq.~(\ref{eq4}) is easily adaptable to computer algebra systems, with built-in $_{p}F_{q}$'s providing improved convergence, see \cite{maple1} for ready-to-use Maple$^{\text{\textregistered}}$ procedure L2S. In Fig.~\ref{fig3} we present $g(\alpha, \beta; x)$ for small values of $\alpha$ and $\beta$ ($\alpha = 1/15 = -\beta$, and $\beta = 1/45$) for which neither Bergstr\"{o}m-Feller formula nor numerical calculations \cite{JPNolan99, RHRimmer05} are applicable.

From the formulae (\ref{eq4}) and (\ref{eq5}) we can retrieve all exactly known cases enumerated in \cite{VVUchaikin99, WRSchneider86, TMGaroni02, AHatzinikitaz08, KAPenson10} and give an unlimited number of new exact solutions $g(\alpha, \beta; x)$; e. g. for $\alpha = 2/3$, $\beta = 1/3$ (here $l = 4$, $k = 6$, $r = 1$, see \cite{ref1}) and $x \geq 0$
\begin{equation}\label{eq6}
g\left(\frac{2}{3}, \frac{1}{3}; x\right) = \sum_{j=1}^{5}\,\frac{c_{j}(4, 6, 1)}{x^{1 + 2 j/3}} \,_{5}F_{6}\left(^{1, \Delta(4, 1 + 2 j/3)}_{\Delta(6, j+1)} \Big\vert \frac{-4^4}{6^6 x^4} \right),
\end{equation} 
and for $x < 0$, where $l = 4$, $k = 6$, $r = 3$:
\begin{equation}\label{eq7}
g\left(\frac{2}{3}, -\frac{1}{3}; x\right) = \sum_{j=1}^{5}\,\frac{c_{j}(4, 6, 3)}{|x|^{1 + 2 j/3}} \,_{5}F_{6}\left(^{1, \Delta(4, 1 + 2 j/3)}_{\Delta(6, j+1)} \Big\vert \frac{4^4}{6^6 |x|^4} \right).
\end{equation} 
These two components neatly match at $x = 0$. The coefficients $c_{j}(4, 6, 1)$, in Eq.~(\ref{eq6}) are equal to $\frac{\Gamma(2/3)}{3\pi}$, $\frac{-2}{9 \Gamma(2/3)}$, $\frac{1}{3\pi}$, $\frac{-5 \Gamma(2/3)}{3^{7/2} \pi}$, $\frac{7\cdot 3^{-11/2}}{\Gamma(2/3)}$, and in Eq.~(\ref{eq7}) $c_{j}(4, 6, 3)$ are equal to $\frac{2\Gamma(2/3)}{3\pi}$, $0$, $\frac{-1}{3\pi}$, $0$, $\frac{14\cdot 3^{-11/2}}{\Gamma(2/3)}$, respectively. This density is depicted in Fig.~\ref{fig4}.

In Fig.~\ref{fig2} we present new distributions $g(7/5, \beta; x)$ for different $\beta$, including symmetric $g(7/5, 0; x)$, not explicitly discussed in \cite{TMGaroni02}. In Fig.~\ref{fig4} we display the PDFs $g(\alpha, 1/3; x)$ for different $\alpha$, including $\alpha = 2/3$, see Eqs.~(\ref{eq6}) and (\ref{eq7}), as well as $g(1/3, 1/3; x)$ confined to $-\infty < x \leq 0$. Our method confirms all the results for symmetric case $\beta = 0$ obtained in related works \cite{AHatzinikitaz08, TMGaroni02}, in which unfortunately no graphical analysis was attempted. Our graphical representations warrant that it is $\alpha$ which grossly determines the global shape as well as their heavy tails. From Fig.~\ref{fig2} we also conclude that $\beta$ has a much weaker influence on their shape as $g(\alpha=\text{const}, \beta; x)$ for different $\beta$ are loosely evocative of each other.

The precise knowledge of distribution functions $g(\alpha, \beta; x)$ is a prerequisite to develop all the theories of anomalous diffusion based on the Fokker-Planck equations \cite{EBarkai01} in conventional and fractional derivatives versions \cite{IMSokolov00, EBarkai01} The new solutions presented here offer a convenient starting point to carry out a systematic study of this approach, as they explicitly contain the parameter $\beta$ (usually set to $\beta = -\alpha$ in earlier attempts). Our solutions will also directly apply to analysis of hydrogen diffusion in the amorphous, high-temperature phase of Pd$_{85}$Si$_{15}$H$_{7.5}$ \cite{RKutner99} in which $\alpha = 1.54$. In econophysical context \cite{EScalas07} the values $\alpha \approx 1.42 - 1.81$ were observed, whereas in \cite{SBorak05} the values $\alpha = 1.64$, $1.78$ and $1.33$ were attributed to the fits of statistics of 2000 Dow Jones Industrial Averages, to 1635 Boeing stock price returns and to the fluctuations of Yen/US\textdollaroldstyle$\,$ exchange rate (1978-1991), respectively.  

In conclusion, the presence of two natural parameters $\alpha$ and $\beta$ in our solutions could permit a precise description of experimental and statistical data characterized by L\'{e}vy distributions. The index $\alpha$ governs the heavy tails whereas $\alpha$ and $\beta$ adjust the position of distribution peaks. We expect that these solutions will be of use in the wide, boundary-crossing field of applications of L\'{e}vy laws.
\ \\

The authors acknowledge support from Agence Nationale de la Recherche (Paris, France) under Program PHYSCOMB No. ANR-08-BLAN-0243-2.


\begin{thebibliography}{10}
\bibitem{ref1} In order to account for possible $\alpha$ and $\beta$, $l$ and $k$ need not be relatively prime natural numbers.
\end{thebibliography}
\end{document}